\documentclass[a4paper,fleqn,usenatbib,useAMS]{mnras}

\usepackage[T1]{fontenc}
\usepackage{ae,aecompl}

\usepackage{graphicx}	
\usepackage{amsmath}	
\usepackage{amssymb}	
\usepackage{color}
\usepackage{graphicx}
\usepackage[compatibility=false]{caption}
\usepackage{subcaption}
\usepackage{dcolumn}
\usepackage{bm}

\title[Non-nested models comparison]{A method for comparing non-nested models \\ with application to astrophysical searches for new physics
}

\author[S. Algeri et al.]{
Sara Algeri$^{1,2}$\thanks{E-mail: s.algeri14@imperial.ac.uk}
Jan Conrad$^{2,3,4,1}$
and David A. van Dyk$^{1}$
\\
$^{1}$Statistics Section, Department of Mathematics, Imperial College London, South Kensington Campus, London SW7 2AZ, United Kingdom\\
$^{2}$Department of Physics, Stockholm University, AlbaNova, SE-106 91 Stockholm, Sweden\\
$^{3}$The Oskar Klein Centre for Cosmoparticle Physics, AlbaNova, SE-106 91 Stockholm, Sweden\\
$^{4}$Wallenberg Academy Fellow
}

\date{Accepted 2016 February 10. Received 2016 February 10; in original form 2015 December 25.}

\pubyear{2016}

\begin{document}
\label{firstpage}
\pagerange{\pageref{firstpage}--\pageref{lastpage}}
\maketitle

\begin{abstract}
Searches for unknown physics and decisions between competing astrophysical models to explain data both rely on statistical hypothesis testing. The usual approach in searches for new physical phenomena is based on the statistical  Likelihood Ratio Test (LRT) and its asymptotic properties. In the common situation,  when neither of the two models under comparison is a special case of the other  i.e., when the hypotheses are non-nested, this test is not applicable. In astrophysics, this problem occurs when two models that reside in different parameter spaces are to be compared. An important example  is  the recently reported excess emission in astrophysical $\gamma$-rays and the question whether its origin is known astrophysics or dark matter.  We develop and study a new, simple, generally applicable,  frequentist method and validate its statistical properties using a suite of simulations studies. We exemplify it on realistic simulated data of the Fermi-LAT $\gamma$-ray satellite, where non-nested hypotheses testing appears in the search for particle dark matter.  

\end{abstract}





This is a pre-copyedited, author-produced PDF of an article accepted for publication in MNRAS Letters following peer review. The version of record \emph{A method for comparing non-nested models with application to astrophysical searches for new physics}, doi: 10.1093/mnrasl/slw025,  is available online at: \url{http://mnrasl.oxfordjournals.org/cgi/content/abstract/slw025?
ijkey=UY4DKz87GlCpToU&keytype=ref}.

\section{Model comparison in astroparticle physics}
\label{sec1}
In astrophysics, hypothesis testing is ubiquitous, because progress is made by comparing competing models to experimental data.  
In the special case, where new physical phenomena are searched for, the most common choice of hypothesis test
is the Likelihood Ratio Test (LRT), whose popularity is partly motivated by the fact that,  assuming $H_0$ is true, the asymptotic distribution  of the LRT statistic is a $\chi^2$. Such result holds  if the regularity conditions specified in Wilks's theorem hold \citep{wilks}.  A key necessary condition is ``nested-ness", meaning that   there is a full model of which both the models under $H_0$ and the alternative hypothesis, $H_1$, are special cases. This is obviously the case for the search for new particles where the null hypothesis (or baseline model), $H_0$, is given by ``background" and $H_1$ is given by ``background+signal of new particle". However, cases where model comparison is non-nested are common: for instance,  when a known astrophysical signal can be confused with new physics, see \cite{ref1} for an example from astroparticle physics, or if the models to be compared reside in different parameter spaces \citep{Profumo}; as in  gamma-ray bursts \citep{sef2}. In these situations, Monte Carlo simulations of the measurement process are often the only possibility, but are challenged by stringent significance requirements, e.g., at the $5\sigma$ level. We present a solution that allows evaluation of accurate statistical significances for non-nested model comparison while avoiding extensive Monte Carlo simulations.  As a concrete example, we apply the proposed procedure to the search for particle dark matter, where the method has particular importance. 

One way to search for dark matter is to consider its hypothesized annihilation products, i.e., $\gamma$-rays,
that can be detected by space borne or ground based $\gamma$-rays telescopes \citep{conradTAMU}.
Here, the issue of source confusion is one of the most challenging aspects of claiming discovery of a dark matter induced signal. A detected excess of $\gamma$-rays may either originate from dark matter annihilation or be caused by conventional, known astrophysical sources. Discrimination can be performed using their spectral distributions, however these are not necessarily part of the same parameter space (see below). This situation arises for example in the search for dark matter sources among the unidentified sources found by Fermi-LAT \citep{ref1}, the claimed detection of a signal consistent with dark matter in our own galaxy, which has gained much attention recently \citep{daylan}, or (once a detection has been made) in  the search for dark matter in dwarf galaxies \citep{PRL1, PRD, PRL3, PRL2, PRL4}. In the recent claims, the existence of a source of $\gamma$-rays (over some background) is established by a LRT, but the crucial and unsolved question is not whether a $\gamma$-ray source exists, but whether it can be explained by conventional sources of $\gamma$-rays as opposed to dark matter annihilation. This is a prime example of an non-nested model comparison.
For definiteness, we can assume  $f(y,E_0,\phi)\propto\phi E_0^{\phi}y^{-(\phi+1)}$ is the probability density function (pdf) 
of  the $\gamma$-rays energies, denoted by $y$, originating from known cosmic sources  and $g(y,M_{\chi})\propto0.73\bigl(\frac{y}{M_{\chi}}\bigl)^{-1.5}\exp\bigl\{-7.8\frac{y}{M_{\chi}}\bigl\}$  is the pdf of the $\gamma$-ray energies of dark matter \citep{bergstrom}. The goal is to decide if $f(y, E_0,\phi)$ is sufficient to explain the data ($H_0$)  or if $g(y,M_{\chi})$ ($H_1$) provides a better fit. 

Although the issue of comparing non-nested models has been addressed since the early days of modern statistics \citep{cox61,cox62,cox13,atkinson,quandt}, as well as in the more recent physical literature \citep{PL05, PL06}, a method with the desired statistical properties, easy implementation and computational efficiency in astrophysics  is still lacking.


This article is arranged as follows. Section~\ref{sec2} reviews  the LRT, Wilks's theorem and their extensions to non-regular situations. Our proposal for testing non-tested models is introduced in Section~\ref{sec3}, validated via simulation studies in Section~\ref{sec4}, and applied to a realistic simulation of the Fermi-LAT $\gamma$-ray satellite in Section~\ref{sec5}. General discussion appears in Section~\ref{sec6}.

\section{Wilks, Chernoff and Trial Factors}
\label{sec2}
Let $f(y;\bm{\alpha})$ and $g(y,\beta)$ be  pdfs of the background and signal, where $y$ is the detected energy, $\bm{\alpha}$ and $\beta$ are parameters.
Suppose observed particles are a mixture of background and source,  i.e.,
\begin{equation}
\label{m1}
(1-\eta) f(y,\bm{\alpha})+\eta g(y,\beta)
\end{equation}
where $0\leq\eta\leq 1$ is the proportion of signal  counts. 
A hypothesis test can be specified as $H_0:\eta=\eta_0$ versus $H_1:\eta>\eta_0$, and if $\beta$ is known
the LRT statistic by
\begin{equation}
\label{T}
T(\beta)=-2\log\frac{L(\eta_0,\hat{\bm{\alpha}}_0,\textrm{-}) }{L(\hat{\eta}_1,\hat{\bm{\alpha}}_1,\beta)},
\end{equation}
where $L(\eta,\bm{\alpha},\beta)$ is the likelihood function 
under \eqref{m1}. The numerator and denominator 
of \eqref{T} are the maximum likelihood achievable under $H_0$ and $H_1$, respectively with $\hat{\bm{\alpha}}_0$ being the MLE of $\bm{\alpha}$ under $H_0$ and $\hat{\bm{\alpha}}_1$ and $\hat{\eta}_1$ the MLEs under $H_1$. 
 \citep{wilks} states that when $H_0$ is true and when testing for a one-dimensional parameter (in this case $\eta$), $T(\beta)$  is asymptotically distributed  as a $\chi^2_1$ (the subscript being the degrees of freedom). 
Among the regularity conditions which guarantee this result are:
\begin{enumerate}
\item[RC1.]The models are nested, meaning that there is a full model of which both $H_0$ and $H_1$ are special cases.
\item[RC2.]The set of possible parameters of $H_0$ is on the interior of that for the full model.
\item[RC3.]The full model is identifiable under $H_0$.
\end{enumerate}
Unfortunately in practice, it is common to encounter non-regular problems. 
Notice for example, if $\beta$ is known  but $\eta_0=0$, RC2 does not hold. 
In this case, \cite{chernoff} applies; it generalizes Wilks and states that if $H_0$ is on the boundary of the parameter space, the asymptotic distribution of $T(\beta)$ is an equal mixture of a $\chi^2_1$ and a Dirac delta function at 0, namely $\frac{1}{2}\chi^2_1+\frac{1}{2}\delta(0)$. 

Further,  if $\eta_0=0$ (on the boundary) and $\beta$ is unknown, the model in \eqref{m1} is not identifiable under $H_0$ and  
RC3 fails. This is known in statistics as a test of hypothesis 
where a nuisance parameter is defined only under $H_1$, or ``trial correction" in astrophysical literature. A solution based on theoretical result of \cite{davies87} is proposed by \cite{gv10}. In particular, under $H_0$, $T(\beta)$ is a random process indexed by $\beta$, specifically if RC2 (but not RC3) holds
$\{T(\beta), \beta \in \bf{B} \}$  is asymptotically a $\chi^2_1$-process. 
A natural choice of test statistic  is $\sup_{\beta} T(\beta)$ and  \cite{gv10} provides an approximation in the limit as $c \rightarrow \infty$ for the tail probability $P(\sup_{\beta} T(\beta)>c)$.
Finally, if both RC2 and RC3 fail to hold (e.g., the important case of $\eta_0=0$ with $\beta$ unknown), we show in our Supplementary Material that because $\{T(\beta), \beta \in \bf{B} \}$ is a $\frac{1}{2}\chi^2_1+\frac{1}{2}\delta(0)$ random process, 
\begin{equation}
\label{ourDC}
P(\sup_{\beta} T(\beta)>c)\approx \frac{P(\chi^2_{1}>c)}{2}+E[N(c_0)|H_0]e^{-\frac{c-c_0}{2}}
\end{equation}
where $E[N(c_0)|H_0]$ is the expected number of upcrossings of the $T(\beta)$ process  over the threshold $c_0$ under $H_0$ and $c_0$ is chosen $c_0<<c$. (Details of how to choose $c_0$ are given in \cite{gv10}, where \eqref{ourDC} is also asserted, but without proof.) Although this approximation holds as $c \rightarrow \infty$, 
when $c$ is small, the right hand side of \eqref{ourDC} is an upper bound for $P(\sup_{\beta} T(\beta)>c)$.
Thus, basing inference on \eqref{ourDC} is valid, though perhaps conservative.



\section{Statistical comparison of non-nested models}
\label{sec3}
Suppose we wish to compare two pdfs, $f(y,\bm{\alpha})$ and $g(y,\beta)$, for which RC1 does not apply, that is the two pdfs are not special cases of a full model and do not share a parameter space. Notice that in both $f$ and $g$
free parameters (i.e., $\bm{\alpha}$ and $\beta$ respectively) are present and thus, the problem cannot be reduced to
a test for  simple hypotheses as in \cite{cousins05}, see \cite{cox61} for more details.
We require  $\beta$ to be one dimensional and $\bm{\alpha}$ to lie in the interior of its parameter space. 
The goal is to develop a test of the hypothesis:
\begin{equation}
\label{test}
H_0: f(y,\bm{\alpha})  \text{ versus } H_1: g(y,\beta) 
\end{equation}

Although $f(y,\bm{\alpha})$ and $g(y,\beta)$ are non-nested, we can construct a comprehensive model which includes both as special cases.
There are two reasonable formulations. We encountered the first in  \eqref{m1}; the second is proportional to $\{f(y,\bm{\alpha})\}^{1-\eta} \{g(y,\beta)\}^{\eta}$, with $0\leq\eta\leq1$ in both formulations. As discussed in \cite{cox62,cox13,atkinson} and \cite{quandt}, 
there are advantages and disadvantages to both. 
From our perspective, the additive form in \eqref{m1} has the advantage of more appealing mathematical properties. 
Since no normalizing constant is involved, the maximization of the log-likelihood reduces to numerical optimization.
In contrast to the test discussed in Section \ref{sec2}, the model in \eqref{m1} is not viewed as a mixture of astrophysical models in which a certain proportion of events, $\eta$, originates a process represented by one model, and the the remaining proportion, $1-\eta$, originates from the completing process represented by the other model. Instead, \eqref{m1} is a mathematical formalization used to embed the pdfs $f(y, \alpha)$ and $g(y,\beta)$ and their corresponding parameters spaces into an overarching model via the auxiliary parameter $\eta$ \citep{quandt}. The overarching model has not astrophysical interpretation, but helps us reformulate the test in \eqref{test} into a suitable form, i.e.,
\begin{equation}
\label{test2}
H_0: \eta=0 \quad \text{versus} \quad H_1: \eta>0.
\end{equation}

Perhaps a more natural formulation of \eqref{test} would be $H_0: \eta=0$  versus $H_1:\eta=1$. 
Unfortunately, neither Wilks's or Chernoff's theorems apply to this formulation since they rely on the asymptotic normality of the MLE under $H_0$, which can only hold if there is a continuum of possible values of $\eta$ under $H_1$,
with $\eta=0$ in its interior.  With indirect dark matter detection, 
the formulation in \eqref{test2} allows the alternative model to include 
both the case where  dark matter and known cosmic sources are present simoultaneously ($0<\eta<1$) and
the case where only dark matter  is present ($\eta=1$).
In situations where intermediate values of $\eta$ are not physical we might, in addition to \eqref{test2}, test $H_0:\eta=1$ versus $H_1:\eta<1$, i.e., 
 interchange the roles of the 
 hypotheses as discussed in \cite{cox62, cox13}. In this case, the nuisance parameter $\bm{\alpha}$ is required to be one dimensional i.e., $\bm{\alpha}=\alpha$.

Under  model  \eqref{m1}, testing \eqref{test2} is equivalent to testing $\eta$ 
on the boundary with  $\beta$ only being defined under the alternative. We
can apply the methods  discussed in Section~\ref{sec2} to solve this problem. 
Notice that such methods can still be applied if the two models share  additional  parameters, $\bm{\gamma}$, i.e., $f(y,\bm{\gamma},\bm{\alpha})$ and $g(y,\bm{\gamma},\beta)$. However,  the maximized likelihoods in \eqref{T} must be replaced by their profile counterparts $L(0,\hat{\bm{\gamma}}_0,\hat{\bm{\alpha}}_0)$ and $L(\bm{\eta}_1,\hat{\bm{\gamma}}_1,\hat{\bm{\alpha}}_1,\beta)$ \citep{davison}.

\begin{figure}
      \vspace{-3\baselineskip}
\centering
        \begin{subfigure}[b]{0.49\textwidth}
                \centering
                \includegraphics[width=\linewidth]{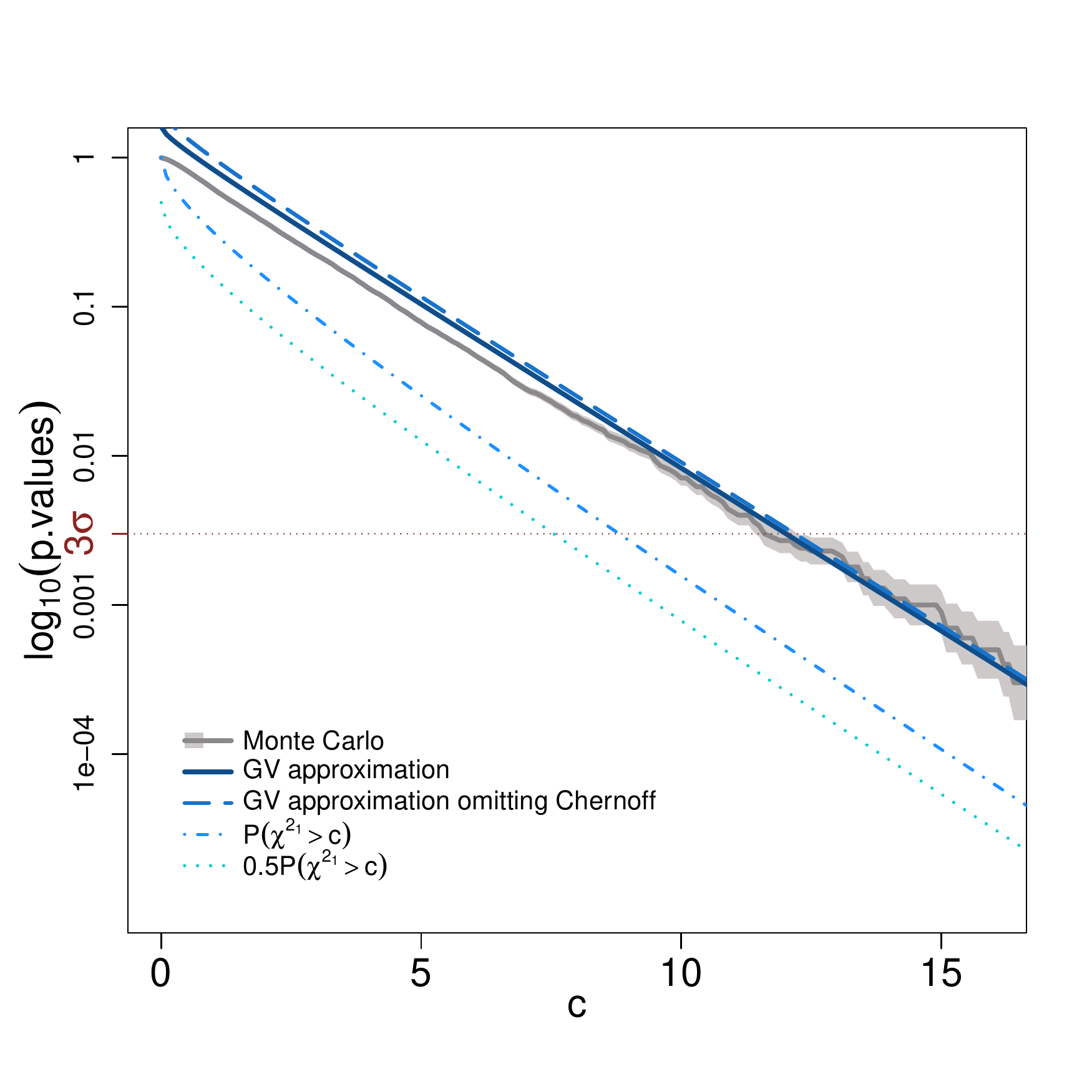}
        \end{subfigure}
\par
\kern\dimexpr-\baselineskip-\parskip-5mm\relax
        \begin{subfigure}[b]{0.49\textwidth}
                \centering
                \includegraphics[width=\linewidth]{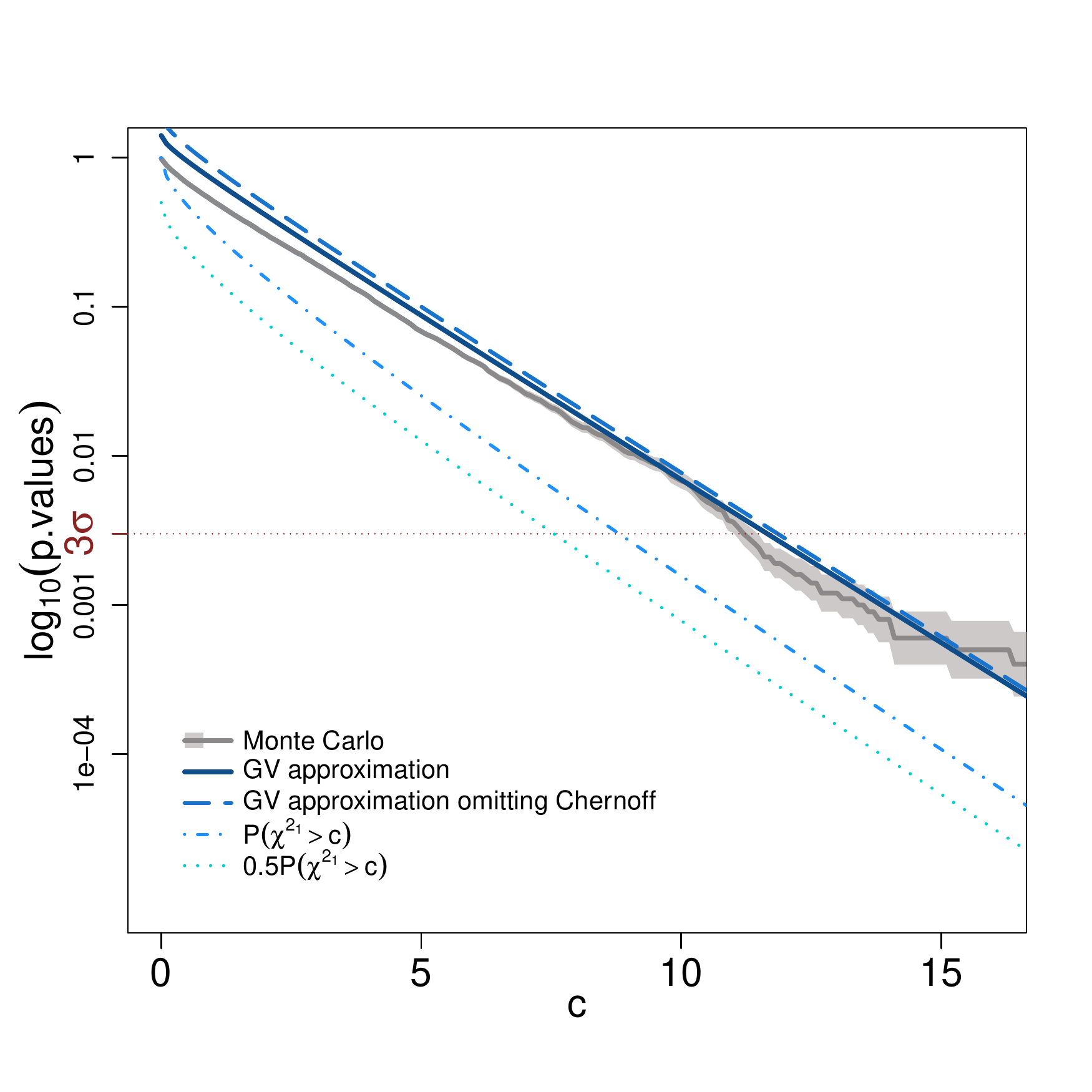}
        \end{subfigure}
      \vspace{-3\baselineskip}
        \caption{
Comparing the approximation in \eqref{ourDC} (solid blue lines) with Monte Carlo estimation of $P(\sup T(M_{\chi})>c)$ (gray dashed lines), for Test 1 (upper panel) and Test 2 (lower panel). Approximations correspondig to \eqref{ourDC} without the Chernoff correction (blue dashed lines), a $\chi^2$ approximation (light blue dash-dotted lines) and a Chernoff-adjusted $\chi^2$ approximation (light blue dotted lines) are also reported.
Monte Carlo p-values were obtained by simulating 10,000 datasets under $H_0$, each of size 10,000 for both simulations. For each simulated dataset $\sup_{M_{\chi}} T(M_{\chi})$ was computed over an $M_{\chi}$ grid  of size $100$ 
for Test 1 and size $400$ for Test 2. Monte Carlo errors (gray areas) were attained via error propagation \citep{cowan}.}
\label{sim}
\end{figure}\begin{figure}
      \vspace{-3\baselineskip}
\centering
        \begin{subfigure}[b]{0.49\textwidth}
                \centering
                \includegraphics[width=\linewidth]{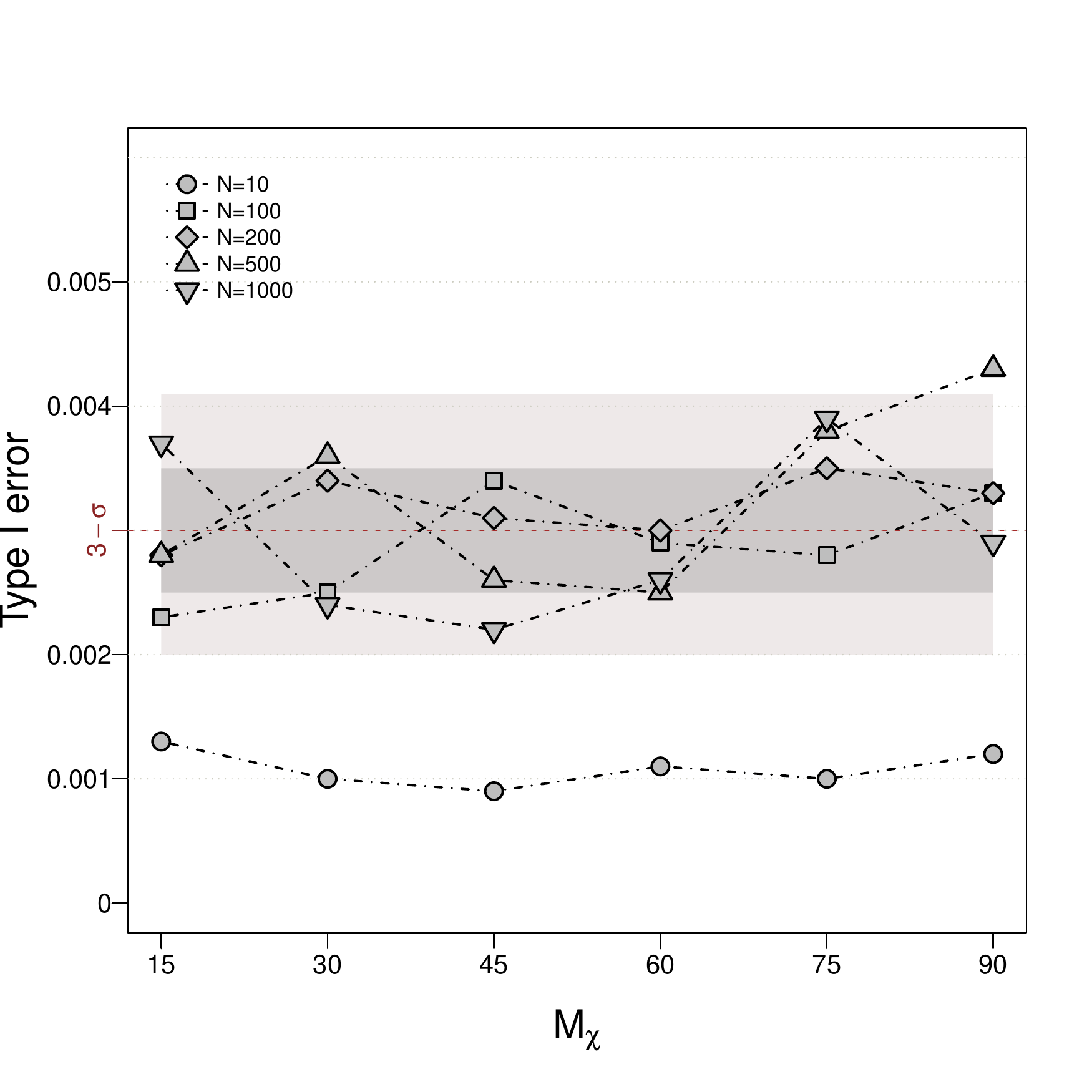}
        \end{subfigure}
\par
\kern\dimexpr-\baselineskip-\parskip-5mm\relax
        \begin{subfigure}[b]{0.49\textwidth}
                \centering
                \includegraphics[width=\linewidth]{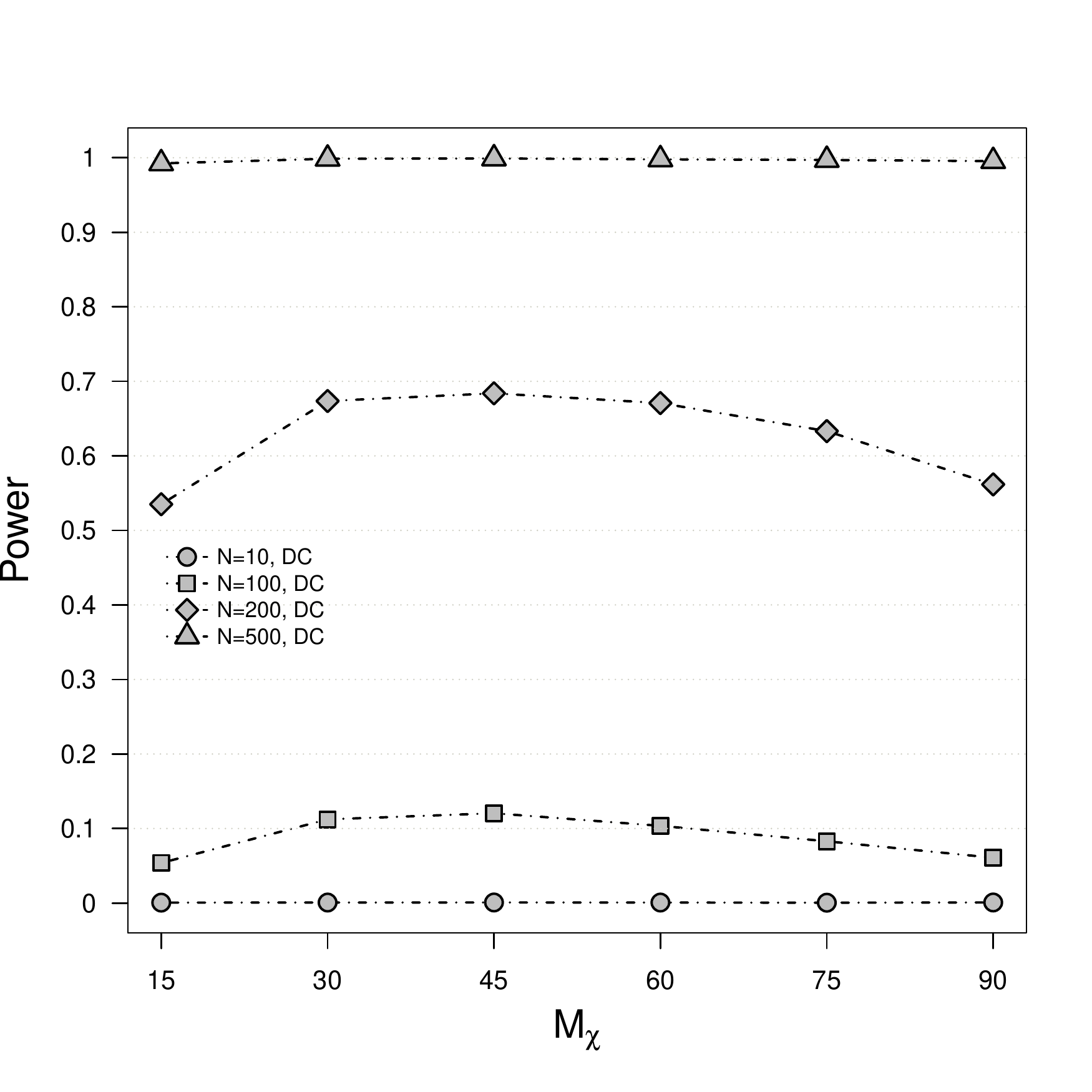}
        \end{subfigure}
      \vspace{-3\baselineskip}
        \caption{Simulated type I errors (upper panel) and power functions (lower panel) for Test 1 with at $3\sigma$ significance. Shaded areas indicate regions expected to contain 68\% (dark gray) and  95\% (light gray) of the symbols if the nominal type I error of 0.003 holds. For both the type I error and power curves 10000 Monte Carlo simulations were used.
}
\label{poweralpha}
\end{figure}
\section{VALIDATION ON DARK MATTER MODELS}
\label{sec4}
We illustrate the reliability of the method proposed for testing non-nested models using two sets of Monte Carlo simulations.
In Test 1, we compare the two models introduced in Section \ref{sec1} with the aim of distinguishing between a dark matter signal and a power law distributed cosmic source.
In Test 2, we make the same comparison but in the presence of power law distributed background.
In this case, $H_0$ specifies as 
\begin{small}
\begin{equation}
\label{f}
f(y,\delta,\lambda,E_0,\phi)= (1-\lambda)\frac{\delta E_0^{\delta}}{k_{\delta}y^{\delta+1}}+\lambda\frac{\phi}{k_{\phi}y^{\phi+1}}E_0^{\phi}
\end{equation}
\end{small}
and  $H_1$ specifies
\begin{small}
\begin{equation}
\label{g}
g(y,\delta,\lambda,E_0,M_{\chi})= (1-\lambda)\frac{\delta E_0^{\delta}}{k_{\delta}y^{\delta+1}}+\lambda\frac{e^{-7.8\frac{y}{M_{\chi}}}}{y^{1.5} k_{M_{\chi}}};
\end{equation}
\end{small}

\noindent
where $k_{\phi}$, $k_{\delta}$ and $k_{M_{\chi}}$ 
are the normalizing constants for each pdf, $0<\lambda<1$, $\delta>0$, $\phi>0$, $E_0=1$, $y\in[E_0,100]$ and $M_{\chi}\in[E_0,100]$.  Note that in this case, the formulation in \eqref{m1}, with mixture parameter $\lambda$, is first used to specify the signal existence over a (relatively well known) background,  whilist in the next step, equation \eqref{m1} is adopted as a merely mathematical tool to treat the non-nested case (as described previously).

For simplicity, in Test 2, $\lambda$, the proportion of events coming from dark matter, was fixed to 0.2. In both tests, we estimated the average number of uprcrossings $E[N(c_0)|H_0]$ using 1,000 Monte Carlo simulations. 
Finally, the approximation to $P(\sup_{\beta} T(\beta)>c)$ is calculated using \eqref{ourDC} on a grid of values
of $c$. The results are compared with the respective Monte Carlo p-values  in Figure \ref{sim} along with the $\chi^2$ and Chernoff corrections one might compute ignoring the regularity conditions in Section \ref{sec2}. 

For small $c$, the approximation in \eqref{ourDC} is greater than its Monte Carlo counterpart. As $c$ increases, however, the approximation converges to the Monte Carlo estimates for a good approximation to the p-value, $P(\sup_{\beta} T(\beta)>c)$.  The $\chi^2$ and respective Chernoff-adjusted approximation lead to over optimistic p-values, whereas similar results to those attained with \eqref{ourDC} are achieved when the factor of 2 that accounts for RC1 is omitted. This is not surprising since the right hand side of \eqref{ourDC}, is dominated by $E[N(c_0)|H_0]$ (which also explains the wide discrepancy between \eqref{ourDC} and the $\chi^2$ approximations in Figure \ref{sim}) and in practice, when testing on the boundary of the parameter space, $E[N(c_0)|H_0]$ is typically calculated simulating  a $\frac{1}{2}\chi^2_1+\frac{1}{2}\delta(0)$ random process directly. Thus, the Chernoff correction is automatically implemented in the leading term of \eqref{ourDC}. 

It is not uncommon in practice, e.g. in astronomy, for the number of counts to be considerably smaller than the 10,000 used in Figure \ref{sim}.
Thus, we conduct a simulation study to verify the type I error (i.e., the rate of false rejections of $H_0$) of the method with smaller samples and verify that the approximate p-value in \eqref{ourDC} holds.
The upper panel of Figure \ref{poweralpha} reports the simulated type I errors  with a detection threshold  on the p-value of 0.003 ($3\sigma$) for different sample sizes when conducting Test 1. 
For sample sizes of at least 100, the Monte Carlo results are consistent with the numerical $3\sigma$ error rate.  
The lower panel of Figure \ref{poweralpha} shows the power (probability of detection) curves at $3\sigma$ of the same test  for different sample sizes. For all the values of 
$M_{\chi}$ considered, a sample size of 500 is sufficient to achieve a power of nearly 1. 

\section{APPLICATION TO SIMULATED DATA FROM the FERMI-LAT}
\label{sec5}
The  Fermi Large Area Telescope (LAT) \citep{fermi} is a pair-conversion $\gamma$-ray telescope on board the earth-orbiting Fermi satellite. 
It measures energies and images $\gamma$-rays between about a 100 MeV and several TeV. One particular aspect is the $\gamma$-ray signal induced by dark matter annihilations, which gives rise to measurable signal from celestial objects, like the Milky Way center or dwarf galaxies.  Here we apply the method proposed in this letter to a dataset simulated with realistic representations of the effects of the detector and present backgrounds. We considered a 5 years observation of putative  dark matter source (dwarf galaxy-like) with dark matter annihilating into  b-quark pairs and a mass of the dark matter particle of 35 GeV. This assumption is consistent with the most generic and popular models for dark matter, namely that it is in large part made of a Weakly Interacting Massive Particles (WIMP). 
 It is also consistent  with recent claims of evidence for dark matter. The signal normalization corresponds to  about 200 events detected in the LAT. Roughly, this corresponds to a dark matter source at the distance of the dwarf galaxy Segue1 (and with comparable dark matter density) and an annihilation cross-section  of  $\sim 2\cdot$10$^{-25}$cm$^{3}$s$^{-1}$). We find a $4.198\sigma$ significance in favor of the dark matter model. Scaling the event rate down to 50 (i.e. considering a lower cross-section by a factor of 4 or lower density by a factor of 16) we obtain $2.984\sigma$  significance (result not shown). Adding complexity, we introduce a background, for example $\gamma$-rays introduced by our own Galaxy. We then considered 2176 counts from a power-law distributed background source as  in \eqref{f}-\eqref{g} and about 550 dark matter events. For simplicity, the  mixture parameter $\lambda$ is fixed at 0.2. In this case, we find $2.9\sigma$  significance in favor of the model in \eqref{g}.  As expected, introducing background significantly reduces the power for distinguishing a dark matter source from a conventional source. It should be noted however that (unlike in a full analysis) we do not attempt to reduce background by taking $\gamma$-ray directions into account.

\begin{table}
 \centering
\begin{tabular}{|c|c|c|c|c|c|c|c|}
\hline
&$H_0$  & N &  $\hat{\eta}$ &    $\hat{M}_{\chi}$ &  $\sup LRT$ & Sig.\\
\hline
Test 1 &  $\eta=0$& 200 & 0.971& 27 & 21.018 & $ 4.038\sigma$\\
\hline
 &  $\eta=1$& 200 & \multicolumn{4}{l|}{p-value $=0.528$}\\
\hline
Test 2   & $\eta=0$&  2726 & 0.999& 30 &12.096&  $2.673\sigma$ \\
\hline
 &  $\eta=1$& 2726 & \multicolumn{4}{l|}{p-value $=1$}\\
\hline
\end{tabular}
\caption{Summary of the analysis on the Fermi LAT simulation comparing the models in Tests 1 and 2. Estimates and Significances refer to the tests $H_0:\eta=0 \quad \text{versus} \quad H_1:\eta>0$. P-values refer to the tests $H_0:\eta=1 \quad \text{versus} \quad H_1:\eta<1$.}
\label{table:realdata}
\end{table}

\section{Summary \& Discussion}
\label{sec6}
We have presented a two-step solution to a common problem in experimental astrophysics: comparing competing non-nested models.
On the basis of the seminal work of  \cite{cox62, cox13} and \cite{atkinson} the first step of our strategy requires the specification of a comprehensive model which extends the parameter space of the models to be compared.  The problem of testing non-nested model is then reduced to the look-elsewhere effect, and thus the second step naturally recalls  \cite{gv10} as an  efficient solution to accurately approximate the significance of new signals. The resulting procedure is easy to implement, does not require extensive calculations on a case-by-case basis and is computationally more efficient than Monte Carlo simulations. Recent developments \citep{usPL} in the nested case illustrate additional desirable statistical properties of \cite{gv10} with respect to \cite{PL05} and \cite{PL06}. Given the nature of the methodology proposed in this letter, we expect these finding to carry over to the non-nested case. 

An example of testing non-nested models arises in the search for particle dark matter. We use this example to validate and illustrate the procedure. We also demonstrate good performance
in a realistic simulation of data that is collected with the Fermi-LAT $\gamma$-ray detector and used in the search for signal from dark matter annihilation.  

Although any pair of hypothesized models can be expressed as a special case of \eqref{m1}, this formulation alone does not always provide a mechanism for a statistical hypothesis test. In the non-nested scenario analysed in this letter, valid inference for the test in \eqref{test2} can be achieved by applying   the methodology in \cite{gv10}. This method, however, cannot handle multi-dimensional parameters that are defined only under $H_1$, nor can it deal with nuisance parameters under $H_0$ which lie on the boundary of their parameter space. A possible approach to tackle the first limitation is to apply the theory in \cite{vg11} to the comprehensive model in \eqref{m1}. Whereas an extention of the method to overcome the second limitation could rely on the theory in \cite{self}. In light of this, the methodology proposed is particularly suited to comparisons of non-nested models where these limitations often do not arise.

Software for the methodology illuatsrated in this letter is available at: \url{http://wwwf.imperial.ac.uk/~sa2514/Research.html}.

\section*{Acknowledgements}

The authors acknowledge Brandon Anderson for using tools publicly available from the Fermi LAT Collaboration to simulate Fermi LAT data. JC thanks the support of the Knut and Alice Wallenberg foundation and the Swedish Research Council. DvD acknowledges support from a Wolfson Research Merit Award provided by the British Royal Society and from a Marie-Curie Career Integration Grant provided by the European Commission.












\bsp	
\label{lastpage}
\end{document}